\documentclass[aps,superscriptaddress,nofootinbib,twocolumn]{revtex4-1}

\usepackage{mathrsfs}
\usepackage{amsfonts}
\usepackage{amssymb}
\usepackage{amsmath}
\usepackage{graphicx}
\usepackage[usenames,dvipsnames]{color}
\usepackage[colorlinks=true,citecolor=blue,linkcolor=magenta]{hyperref}
\usepackage{ulem}
\usepackage{lmodern}

\newcommand{\figpath}{./}

\newcommand{\ket}[1]{\vert{ #1 }\rangle}

\begin{document}

\title{Hierarchical surface code for network quantum computing\\ with modules of arbitrary size}

\author{Ying Li}

\affiliation{Department of Materials, University of Oxford, Parks Road, Oxford OX1 3PH, United Kingdom}

\author{Simon C. Benjamin}

\affiliation{Department of Materials, University of Oxford, Parks Road, Oxford OX1 3PH, United Kingdom}

\date{\today}

\begin{abstract}
The network paradigm for quantum computing involves interconnecting many modules to form a scalable machine. Typically it is assumed that the links between modules are prone to noise while operations within modules have significantly higher fidelity. To optimise fault tolerance in such architectures we introduce a hierarchical generalisation of the surface code: a small `patch' of the code exists within each module, and constitutes a single effective qubit of the logic-level surface code. Errors primarily occur in a two-dimensional subspace, i.e.~patch perimeters extruded over time, and the resulting noise threshold for inter-module links can exceed $\sim 10\%$ even in the absence of purification. Increasing the number of qubits within each module decreases the number of qubits necessary for encoding a logical qubit. But this advantage is relatively modest, and broadly speaking a `fine grained' network of small modules containing only $\sim 8$ qubits is competitive in {\it total} qubit count versus a `course' network with modules containing many hundreds of qubits. 
\end{abstract}

\maketitle

\section{Introduction}
\label{sec:Introduction}

There are two different approaches to fabricating a large-scale universal quantum computer. One is to create a single `monolithic' architecture in which each qubit is directly and deterministically connected to its neighbours. An alternative is the network architecture~\cite{Dur2003, Benjamin2006, Campbell2007, Jiang2007, Benjamin2009, Li2012, Fujii2012, Monroe2014}, where a single quantum computer is formed from numerous interlinked small devices, \textit{modules}, each having only a modest number qubits and correspondingly little computational power. This approach may prove to be well suited to ion trap systems~\cite{Monroe2014, Harty2014, Ballance2014} or colour centres in diamond~\cite{Bernien2013}, where optical activity can be directly harnessed to create a photonic link; modules comprised of superconducting qubits can also be networked either via microwave links~\cite{Roch_RemoteEntanglement} or by exploiting microwave to optical converters. It is likely that the size of a module, i.e.~the number of physical qubits within it, may vary dramatically according to the technology: whereas a colour centre might have at most a dozen or so satellite nuclear spins, a superconducting module could easily be envisaged as a grid of hundreds of qubits. It is therefore interesting to ask what impact the module size has on performance characteristics such as the fault tolerance threshold, and thus the total number of physical qubits needed per logical qubit. 

An advantage of the network architecture is its manifest scalability. However, based on experimental results to-date it is reasonable to assume that inter-module communications will only provide low-quality entanglement~\cite{Moehring2007, Bernien2013, Hucul2015} compared with intra-module quantum gates~\cite{Brown2011, Harty2014, Ballance2014, Barends2014}. Whatever approach one adopts to mitigate the noise on the links, there will inevitably be a resource cost versus an idealised monolithic architecture were all gates are of comparable fidelity to the intra-module operations. In other words, to implement the same quantum algorithm, more qubits are required on the network architecture to overcome network noise. A goal of this paper is to quantify this difference. 

\section{Outline of approach}
\label{sec:Outline}

We investigate quantum computing with a network architecture involving modules containing from only two qubits to about a thousand qubits. In our study, we exploit two methods to negate errors: entanglement purification~\cite{Dur2003, Jiang2007} and error correction via the surface code~\cite{Kitaev2003, Dennis2002}. Entanglement purification is a low-level process that corrects errors in inter-module links and is carried out individually within each module with the help of classical communications. We use the term `broker unit' for the dedicated hardware (comprising one or more qubits) associated with entanglement purification. 

For small modules with only a few qubits in total, each module only provides one qubit participating the surface code while the rest are involved in purification. This is equivalent to architectures that have been studied in earlier papers~\cite{nickersonNC2013, benjaminPRX2015}. The challenge we tackle here is to efficiently exploit large modules with at least tens of qubits; our solution retains the purification but additionally introduces a hierarchical variant of the surface code. A piece of surface code (or `patch') exists in every module, such that each module can be effectively regarded as a single qubit in a higher (logical level) surface code. There are interesting consequences for the localisation and correction of errors, given that such errors tend to occur at the boundaries between the modular patches. In essence the errors live in a two-dimensional space, one spatial and one temporal dimension, so that the relevant threshold is {\it equivalent} that of a two spatial dimension system with perfect noise-free stabiliser measurement. In this way the hierarchical surface code tolerates network errors up to $15\%$, and purification techniques need only bring the network noise within this limit. 

One might expect that the performance of a network computer will approach that of a monolithic computer as the module size increases. To compare the resource requirements we study the number of qubits required for encoding a single logical qubit. We find that the total number of physical qubits required per logical qubit does indeed decrease with the size of modules. For a practical level of network noise and modules containing hundreds of qubits, the cost of encoding a logical qubit on the network architecture is still about nine times higher than the cost on the monolithic architecture. Meanwhile we find that for a `fine grained' network comprised of small modules containing only $\sim 8$ qubits, the overhead versus the monolithic system is a factor of about fifteen. It is perhaps surprising that the resource cost associated with adopting the flexible network paradigm varies so little over a wide range of module sizes, i.e.~network granularity does not strongly affect the total resource cost. 

% monolithic
% 1521
% simple
% 5     6     7     8     9    10
% 383645       26934       22743       22472       25281       28090
% large
% 13125

\section{System}
\label{sec:System}

\begin{figure}[tbp]
\centering
\includegraphics[width=0.85\linewidth]{\figpath 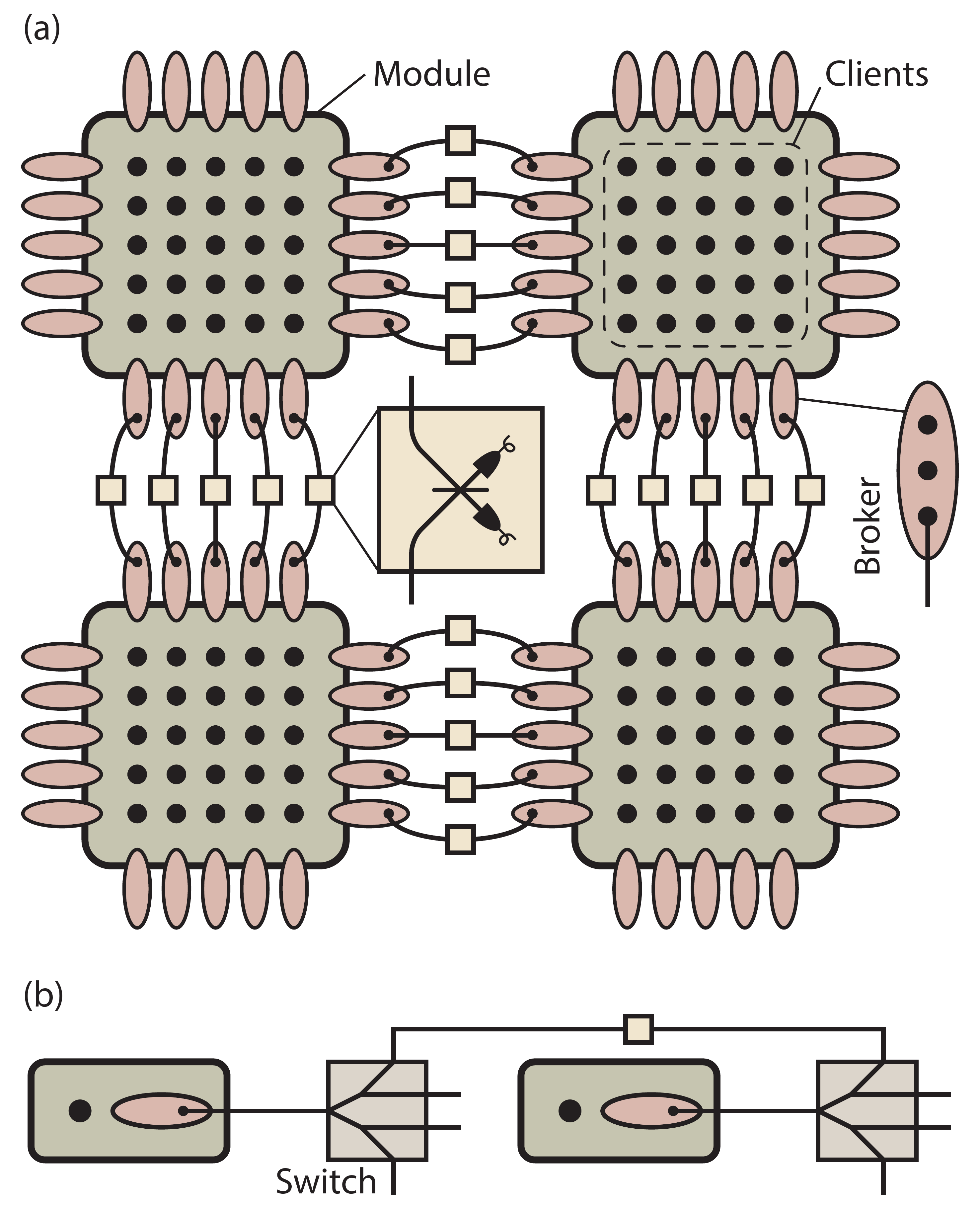}
\caption{
A quantum computer built with optically networked quantum modules. Black circles represent qubits. (a) Each module contains a $D\times D$ client-qubit array as well as $4\times D$ broker units on the perimeter, where $D$ is the dimension of module. The modules in this figure have $D = 5$. Broker units achieve entanglement with one another via inter-module optical couplings. Typically the raw entanglement will be generated by a joint measurement on photons emitted from optically active broker qubits. Inside a broker unit, there are additional ancilla qubits which are used to purify raw entanglement to a higher fidelity. (b) Each simple module only contains one client qubit and one broker unit. A simple module is coupled with four neighbouring modules via a switch for rerouting the optical connection. 
}
\label{fig:module}
\end{figure}

We consider a quantum computer built with networked quantum modules as shown in Fig.~\ref{fig:module}. We focus on the case that each module contains an array of \textit{client} qubits and a series of entanglement-purifying \textit{broker units} on the perimeter of the client array; each broker unit contains several qubits as we presently discuss [see Fig.~\ref{fig:module}~(a)]. Quantum information is stored in client qubits, and brokers are used to generate entanglements between neighbouring modules. In each broker unit, there must be at least one qubit that is optically coupled with another module. Raw entanglement prepared with the optical coupling is purified with the help of other qubits in broker units; the qubits forming the surface code `patch' therefore never `see' the raw entanglement, only the purified form. To provide a context for assessing the performance of the hierarchical surface code, we also consider the limit of small modules where each module may contain only one client qubit and one broker unit [see Fig.~\ref{fig:module}~(b)], in which case the sole broker unit services links to all connected modules by rerouting the optical connection as required. These small modules do not use hierarchical surface code, instead relying on a single surface code layer. 

\begin{figure}[tbp]
\centering
\includegraphics[width=0.9\linewidth]{\figpath 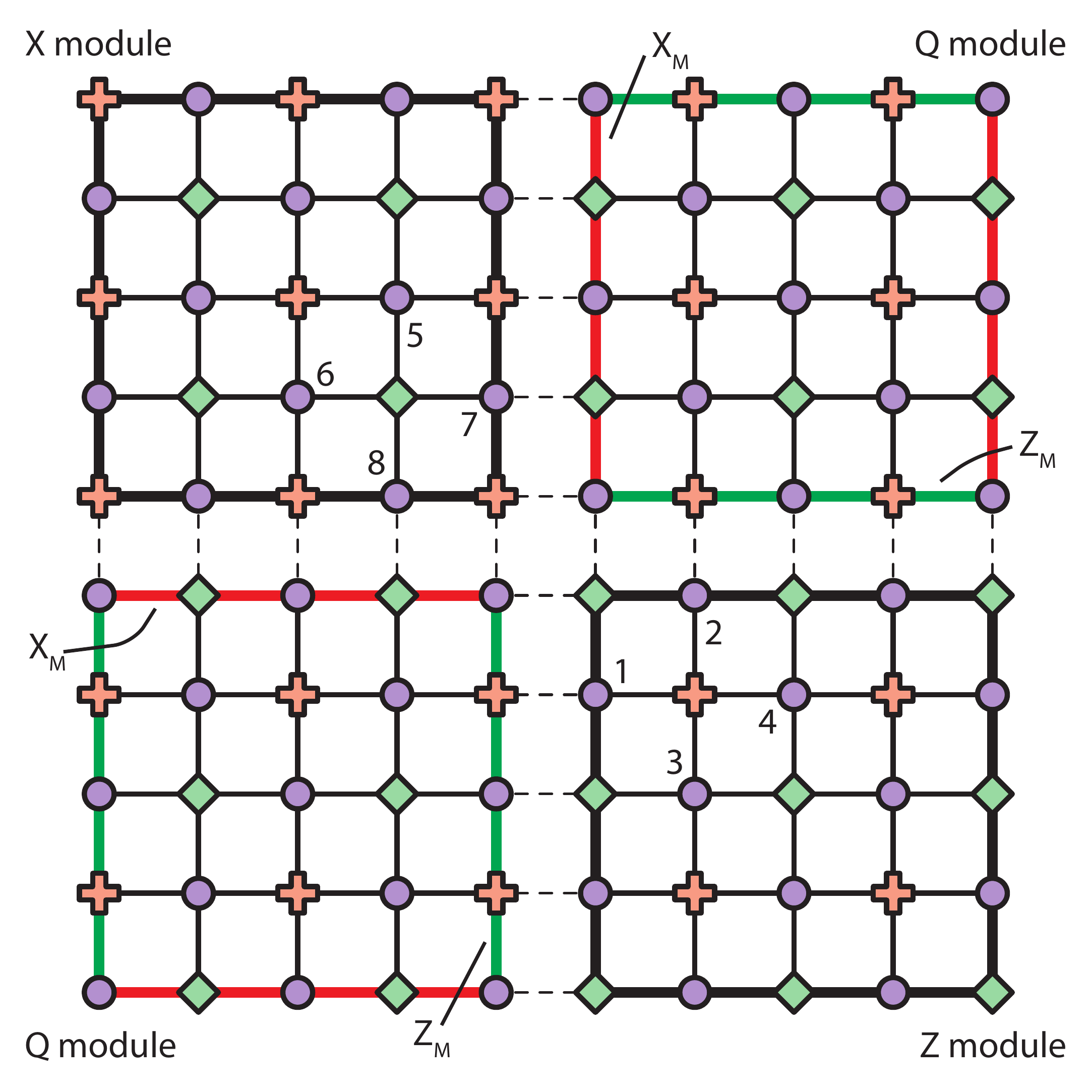}
\caption{
Square lattice of qubits for implementing the surface code. Data qubits, X ancillary qubits and Z ancillary qubits are represented by circles, crosses and squares, respectively. The CNOT gate can be performed on any pair of qubits connected by an edge. The figure depicts four modules each with $D = 5$: solid edges correspond to intra-module gates, and gates of dashed lines are implemented with inter-module entanglement, i.e.~distributed CNOT gates. Errors arising from imperfectly purified remote entanglement will only affect qubits on the perimeter of a module, these are marked with bold edges. $X_\text{M}$ and $Z_\text{M}$ denote Pauli operators of module qubits. 
}
\label{fig:code}
\end{figure}

Intra-module CNOT gates are performed via interactions between qubits within the module. We assume that these CNOT gates are available for any pair of nearest-neighbouring qubits in the same module. For client qubits in different modules, \textit{distributed} CNOT gates are performed by consuming entanglement that has been generated between brokers. The circuit for the distributed CNOT gate~\cite{Eisert2000} is shown in Fig.~\ref{fig:circuit}~(a).

With the geometry of modules in Fig.~\ref{fig:module}, client qubits on the perimeters of neighbouring modules are indirectly coupled through brokers. Therefore, ultimately a square lattice is formed by all client qubits of the network, in which (intra-module or distributed) CNOT gates are available for any pair of nearest-neighbouring qubits. With such a lattice, we can implement the surface code across the entire module network. 

Within the surface-code lattice (Fig.~\ref{fig:code}), qubits are divided into three groups: data qubits (circles), and measurement-enabling qubits of two kinds: X ancillary qubits (crosses) and Z ancillary qubits (squares). The subspace for encoding information in the collective is defined by enforcing sets of stabilisers $XXXX$ and $ZZZZ$, which are products of Pauli operators on four data qubits surrounding X ancillaries and Z ancillaries, respectively~\cite{Kitaev2003,Dennis2002}. Errors are detected by repeatedly measuring stabilisers~\cite{Fowler2009} with circuits shown in Fig.~\ref{fig:circuit}~(d)~and~(e). 

\section{Modular surface code and thresholds}
\label{sec:Thresholds}

\begin{figure}[tbp]
\centering
\includegraphics[width=0.9\linewidth]{\figpath 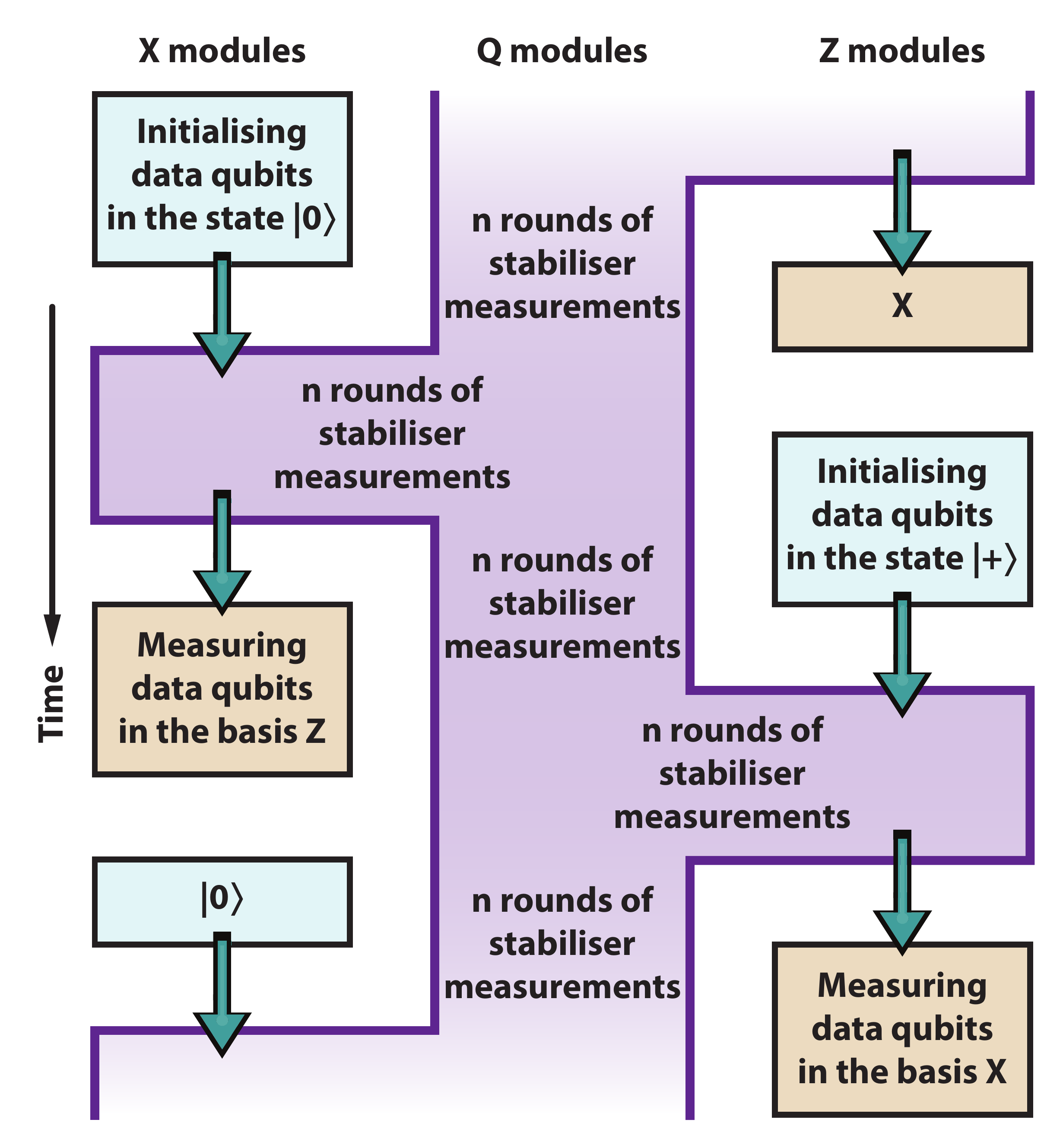}
\caption{
Protocol for module-qubit stabiliser measurements. Each full round of module-qubit stabiliser measurements involves measuring both X stabilisers and Z stabilisers. To measure X (Z) stabilisers of module qubits, data qubits in X (Z) modules are initialised in the state $\ket{0}$ ($\ket{+}$, which is prepared by a Hadamard gate on the state $\ket{0}$) and measured in the $Z$ basis ($X$ basis, which can be realised with the measurement in the $Z$ basis after a Hadamard gate) after $n$ rounds of physical-qubit stabiliser measurements. When measurements of module-qubit X (Z) stabilisers are in progress, Z (X) modules are not involved in physical-qubit stabiliser measurements. Between each set of module-qubit stabiliser measurements, physical-qubit stabiliser measurements are performed on only Q modules for $n$ rounds. 
}
\label{fig:flow}
\end{figure}

In our modular network, errors associated with the entanglement generated over network links are first reduced by entanglement purification. After the purification, there are still some residual errors in the inter-module entanglement because of the limited resources of each broker unit. These residual entanglement errors, together with errors arising from intra-module operations, are finally corrected by the surface code. Assuming the entanglement is ideally in the form $(\ket{00}+\ket{11})/\sqrt{2}$, we model the error-burdened entangled state as
\begin{eqnarray}
\mathcal{E} = F[\openone] + p_\text{X}[X] + p_\text{Z}[Z] + p_\text{Z}[Z],
\label{eq:EntError}
\end{eqnarray}
where $F = 1-p_\text{X}-p_\text{Y}-p_\text{Z}$ is the fidelity, the superoperator $[U]\rho = U\rho U^\dag$, and $X,Y,Z$ are Pauli operators on one of two entangled qubits. For intra-module operations, we assume a qubit may be initialised in the incorrect state with the probability $\epsilon_\text{I}$; the measurement may report an incorrect outcome with the probability $\epsilon_\text{M}$; and each single-qubit gate and CNOT gate may induce an error with the probability $\epsilon_1$ and $\epsilon_2$, respectively. A noisy gate is modelled as a perfect gate followed by single-qubit depolarizing noise for single-qubit gates and two-qubit depolarizing noise for the CNOT gate~\cite{Raussendorf2007PRL}. 

As one might expect, we find that if we consider modules containing a larger client array then more residual entanglement errors can be corrected with the surface code. This would be true even if we were to simply regard all the `patches' of surface code as part of a single surface without giving any special status to the borders between patches. However, in doing so we would be failing to exploit our knowledge that errors are more common along the perimeters. To properly exploit the potential advantage of large modules, instead of continuously performing stabiliser measurements on the entire surface-code lattice, we introduce an intermediate encoding based on the client array of each module. 

Similar to physical qubits on the surface-code lattice, modules are also divided into three groups: Q modules, X modules and Z modules (see Fig.~\ref{fig:code}). The client array of each Q module itself is a piece of complete surface-code lattice, hence one informational qubit can be encoded in each Q module, and we refer to this as a \textit{module qubit}. Moreover, the X modules and Z modules have roles similar to X ancillary qubits and Z ancillary qubits in the basic surface code: they perform X-stabiliser and Z-stabiliser measurements on neighbouring module qubits, respectively. Therefore, a network of modules forms a surface code on a higher level where each module-qubit now constitutes a basic unit. A logical qubit of the algorithm being executed on the computer will be encoded at this level, thus spanning multiple modules. 

The protocol for performing stabiliser measurements on the module-qubit level is shown in Fig.~\ref{fig:flow}. The stabiliser $XXXX$ of four module qubits equals the product of all physical-qubit stabilisers $XXXX$ within the X module. To measure the X-stabiliser of four module qubits, firstly data qubits in the the corresponding X module are all initialised in the state $\ket{0}$; secondly physical-qubit stabiliser measurements across the X module and measured Q modules (see Fig.~\ref{fig:flowII}) are performed for several rounds; finally all data qubits in the X module are measured in the $Z$ basis (see Fig.~\ref{fig:flow}). The module-qubit Z-stabiliser measurement is analogous. Between each set of module-qubit stabiliser measurements, physical-qubit stabiliser measurements are performed on only Q modules for several rounds. 

Commensurate with our two-tier encoding, the error correction includes two tiers. In the first tier the outcomes of physical-qubit stabiliser measurements are used to correct errors on the physical-qubit level. As we describe below, this process can exploit the highly inhomogeneous nature of the physical errors i.e.~the high error density at the border of each module's `patch' of surface code. Once this step is complete, only sequences of errors (error chains) that span entire module qubits can survive. For Q modules, such an error is a bit or phase flip of that specific module qubit. For the X and Z modules, whose role is to provide stabiliser measurement on the surrounding four Q modules, the consequence of such an error chain is that the stabiliser outcome is incorrectly evaluated (i.e.~it is the inverse of the correct outcome). Both types of errors are handled in the second tier of the process, where one simply regards each Q module as a data qubit, and errors on these qubits are determined by analysis of the imperfect stabiliser measurements in the standard way (regardless of the fact that those measurements derive from entire X and Z modules). 

\begin{figure}[tbp]
\centering
\includegraphics[width=0.9\linewidth]{\figpath 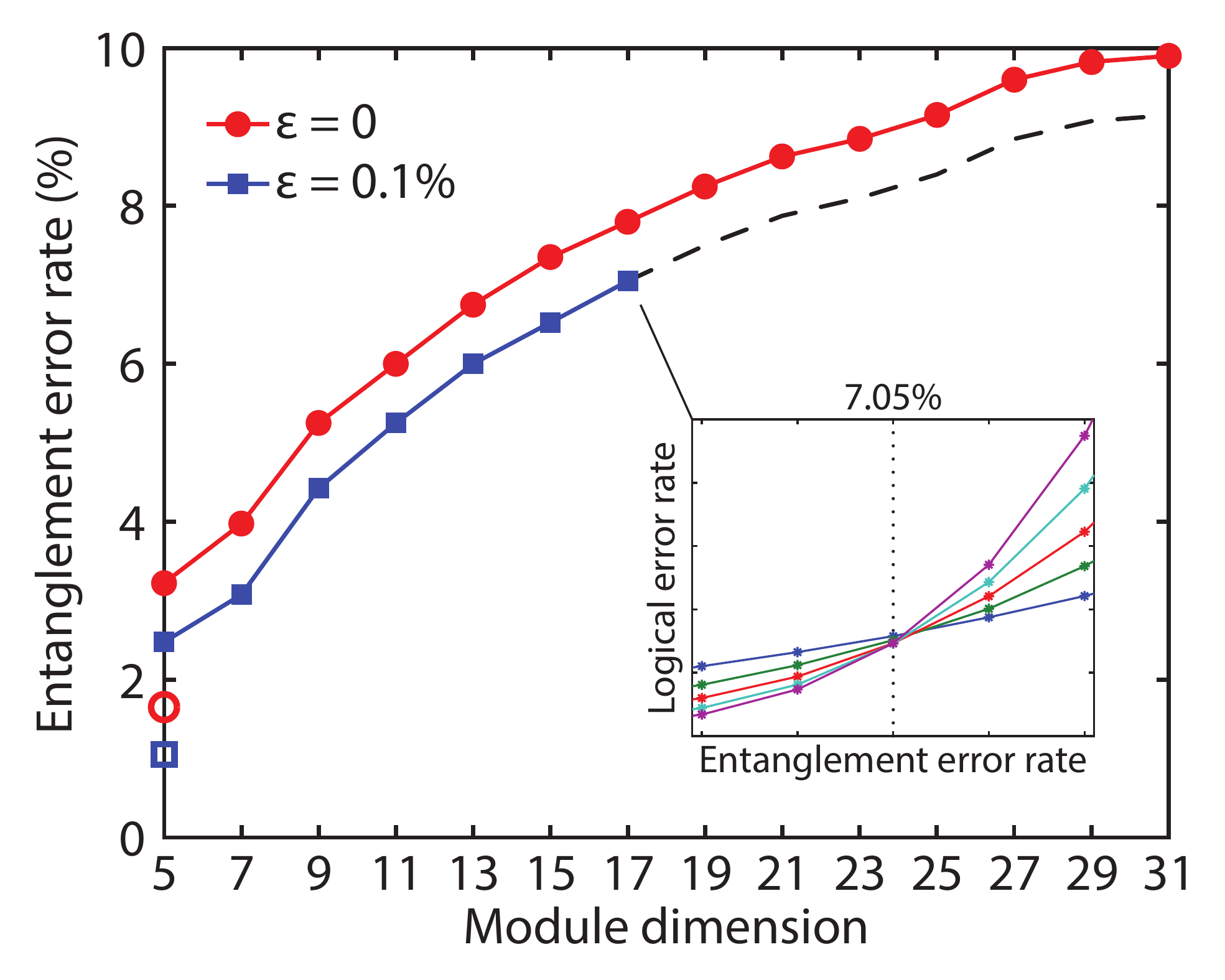}
\caption{
Fault tolerance thresholds in terms of the rate of errors on inter-module entanglement ($1-F$) for modules with different dimensions $D$. Here $F$ is the fidelity {\it subsequent} to any purification within broker units, so that we are seeing the corrective power of the hierarchical surface code alone. For context, thresholds for simple modules are marked with the empty circle and square on the left vertical axis. We have assumed that entanglement errors are unpolarised, i.e.~$p_\text{X}=p_\text{Y}=p_\text{Z}$, and all intra-module operations have the same error rate $\epsilon_\text{I} = \epsilon_\text{M} = \epsilon_1 = \epsilon_2 = \epsilon$. These thresholds are obtained by counting errors on logical qubits encoded in $(2L-1)\times (2L-1)$ module arrays (see Appendix~\ref{sec:AppThreshold} for details). The inset shows logical error rates for $D = 17$ and $L = 3,5,7,9,11$. With an entanglement error rate lower than the threshold (the dotted line in the inset), the logical error rate decreases with $L$ (the left side of the threshold in the inset). The dashed line denotes inferred thresholds according to the difference (which is $\sim 0.75\%$) between two solid lines. 
}
\label{fig:threshold}
\end{figure}

To understand how the inhomogeneity in the distribution of errors is exploited, consider first the artificial case that intra-module operations are perfect (i.e.~$\epsilon_\text{I} = \epsilon_\text{M} = \epsilon_1 = \epsilon_2 = 0$). Then all errors are due to imperfectly-purified inter-module entanglement, and so errors occur strictly on perimeters of client arrays (bold lines in Fig.~\ref{fig:code}). During X-stabiliser measurements of module qubits, in a Q module bit errors and stabiliser-measurement errors occur with the rate $p_\text{X}+p_\text{Y}$ on the two boundaries facing X modules (red bold lines), and in an X module phase errors and stabiliser-measurement errors occur with the rate $p_\text{Z}+p_\text{Y}$ on the entire perimeter (black bold lines). We note that on the corner of X modules, error rates are approximately doubled. It is similar for Z-stabiliser measurements of module qubits. Here, $p_\text{X},p_\text{Y},p_\text{Z}$ are error rates in the inter-module entanglement {\it after} any purification has taken place. Errors are restricted to the one-dimensional perimeter of modules, but the correction process involves $n$ rounds and therefore the syndrome matching occurs in a two-dimensional space: one spatial and one temporal. This is in contrast to a standard surface code approach with homogeneous errors in gates and measurements, where we would need to match syndrome outcomes in a three-dimensional array. There is a very significant advantage in terms of the threshold: whereas the 3D threshold is in the region of $1\%$, for the restricted 2D case it is $10\%$~\cite{Dennis2002}. Therefore, if entanglement error rates satisfy $p_\text{X}+p_\text{Y},p_\text{Z}+p_\text{Y}<10\%$, we are below threshold and thus the rates of module-qubit errors after the first step of error correction decrease with the dimension of modules (i.e.~size of two-dimensional error-correction lattices). Moreover, if indeed the module-qubit error rates decrease with the module dimension, then the threshold for errors in the inter-module entanglement increases with the module dimension. In the limit of large modules, the threshold of entanglement error rate should approach $15\%$ for depolarising errors, i.e.~$p_\text{X}=p_\text{Y}=p_\text{Z}$. If we now allow for a small but finite rate of errors for intra-module operations and measurements, the proceeding remarks all apply except that occasional errors will occur within the perimeter of the `patches' with the consequence that the tolerance of noise on the inter-module links will be somewhat reduced. 

\begin{figure*}[tbp]
\centering
\includegraphics[width=0.90\linewidth]{\figpath 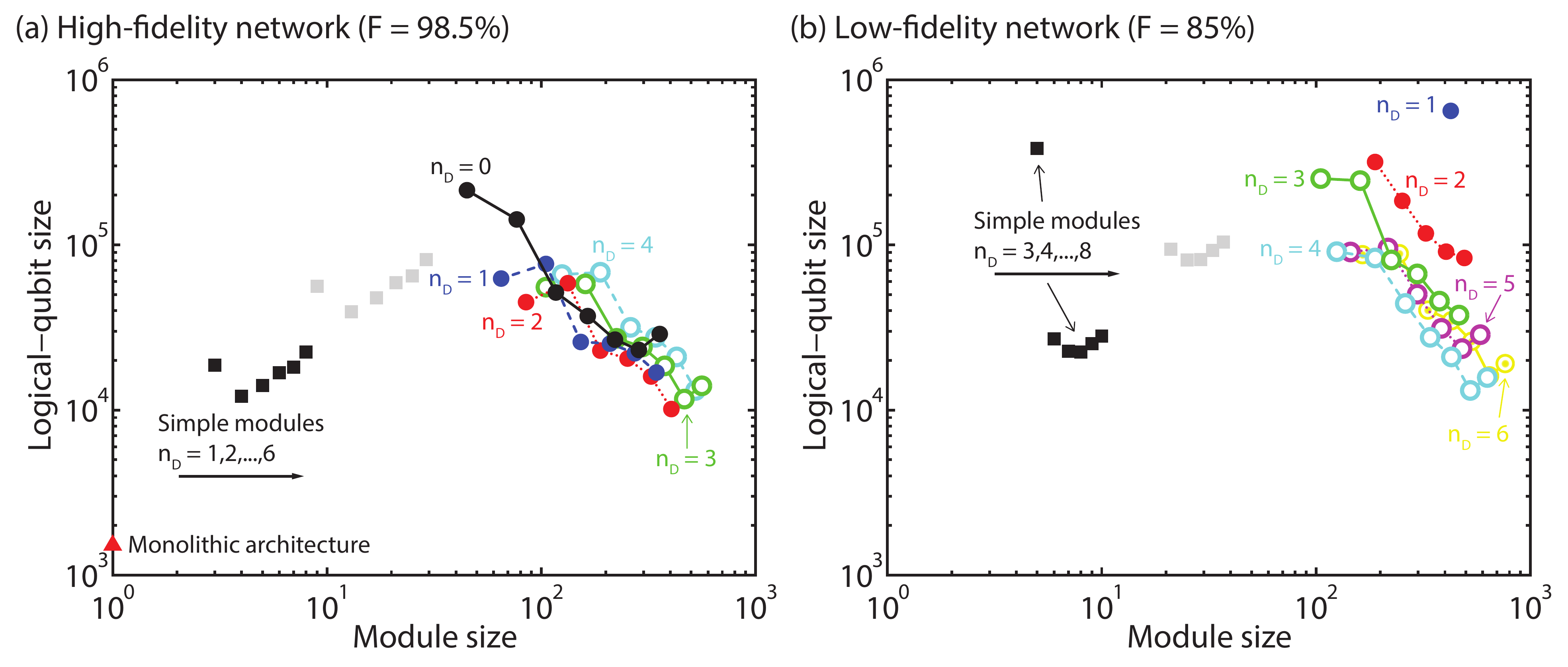}
\caption{
The total count of physical qubits per logical qubit required to achieve the logical error rate $\epsilon_\text{L} = 10^{-12}$. $F$ is the fidelity of the `raw' entanglement between modules; we have assumed uniform raw entanglement errors in the form of Eq.~(\ref{eq:EntError}), i.e.~$p_\text{X}=p_\text{Y}=p_\text{Z}$, and all intra-module operations have the same error rate $\epsilon_\text{I} = \epsilon_\text{M} = \epsilon_1 = \epsilon_2 = 0.1\%$. Circles represent qubit costs for modules with the dimension $D\geq 5$, so that a module qubit can be encoded in each Q module. For such modules, the module size (the total number of qubits in each module) equals $D^2+4D(n_\text{D}+1)$, where each broker contains $n_\text{D}+1$ qubits, and $n_\text{D}$ is number of entanglement purification tiers (see Appendix~\ref{sec:AppTime}). Black squares represent qubit costs for simple modules with only one broker. In each simple module, the qubit number is $n_\text{D}+2$. Gray squares correspond to simple modules with four brokers, i.e.~the module size equals $4n_\text{D}+5$. The red triangle on the left vertical axis of (a) represents the qubit cost on the monolithic architecture~\cite{Fowler2012}, i.e.~when noisy network links are not used and {\it all} gates are performed with the lower error rate of $0.1\%$. These qubit costs are obtained by numerically obtaining parameters $\epsilon_0$ and $\kappa$ in Eq.~(\ref{eq:scaling}) (shown in Fig.~\ref{fig:parameters}, see Appendix~\ref{sec:AppCosts} for details). 
}
\label{fig:cost}
\end{figure*}

In Fig.~\ref{fig:threshold} we show the results of a series of numerical simulations which verify this analysis. The figure shows the fault tolerance threshold for the rate of errors on the inter-module entanglement, assuming that (purified) entanglement errors are uniform over the $X$, $Y$ and $Z$ channels. Note that this is not a favourable assumption: if the errors were not uniform, this would be an opportunity to enhance the threshold by exploiting this knowledge in the decoder. The numerical results reveal that the threshold indeed increases with the module dimension, which coincides with expectations from the preceding analysis. For comparison we also find thresholds for simple modules, i.e.~each module only contains one client qubit as so we do not use the hierarchical approach. 

The observed thresholds vary from $1.65\%$ to $9.9\%$ depending on the size of modules. When we allow for errors induced by intra-module operations, the ability to correct errors on the inter-module operations is reduced as expected, i.e.~the threshold rate of tolerable inter-module errors decreases with the error rate of intra-module operations. Taking all intra-module operations to have the same error rate $\epsilon_\text{I} = \epsilon_\text{M} = \epsilon_1 = \epsilon_2 = 0.1\%$, we find that the threshold of entanglement error rate is reduced by $0.75\%$. As shown in the figure, this reduction varies only very slightly with the module dimension. 

\section{Qubit costs}
\label{sec:Costs}

In the preceding analysis, we considered the structure of the hierarchical surface code and its threshold in terms of the rate of errors on inter-module entanglement; the error rate was taken to be post-purification. Now in order to find the optimal structure for a network architecture, we must consider the power and cost of the brokering units and optimise the number of qubits assigned to that role. We can then find the overall resource cost of fault tolerant computing given a specific error rate on the `raw' inter-module entanglement. 

In a quantum computer based on the surface code, the unit of quantum computing is a logical qubit encoded in a $(2L-1)\times (2L-1)$ qubit (module) array, where the array distance $L$ is the minimum number of data qubits (Q modules) for defining a logical Pauli operator. Given that operations are performed with an error rate lower than the system's threshold, the rate of logical errors decreases with the size of the logical qubit. The logical error rate per surface code cycle, i.e.~a full round of stabiliser measurements, scales with the distance $L$ as~\cite{Fowler2012}
\begin{eqnarray}
\epsilon_\text{L} \simeq \epsilon_0 e^{-\kappa L},
\label{eq:scaling}
\end{eqnarray}
where parameters $\epsilon_0$ and $\kappa$ are determined by error rates of operations. 

In our network architecture, qubits used for entanglement generation and purification do not participate forming logical qubits, i.e.~these qubits assigned to the `broker units' are an additional cost due to the modular architecture. It is non-trivial to optimise the partitioning of qubits between broker units and the internal client arrays, in such a way as to minimise the total number of qubits needed to achieve a given logical error rate. The size of logical qubits (determined by parameters $\epsilon_0$ and $\kappa$) depends on the inter-module entanglement error rate {\it after} purification, and this rate will improve as more qubits are dedicated to purification. 

By numerically obtaining parameters $\epsilon_0$ and $\kappa$, we can find the cost of physical qubits per logical qubit (logical-qubit size). We have considered error rates of intra-module operations $\epsilon_\text{I} = \epsilon_\text{M} = \epsilon_1 = \epsilon_2 = 0.1\%$ and two possible values for the error rate on the \textit{raw} entanglement: $1-F = 1.5\%$ and $15\%$. The logical-qubit size for achieving the logical error rate $\epsilon_\text{L} = 10^{-12}$ is shown in Fig.~\ref{fig:cost}. The `raw' entanglement error rate $1-F = 1.5\%$ is tenth of the theoretical threshold in the large module limit. For this small entanglement error rate, we expect that the purification is not necessary for large modules. However a raw entanglement error rate $1-F = 15\%$ is more practical for current technologies. For such a large error rate, we see that entanglement purification is always necessary. 

In general, the qubit cost decreases with the size of modules (total number of qubits in each module). When the size of modules approaches a thousand qubits, the qubit cost is about nine times higher than the cost on the monolithic architecture. We note that a factor of two is due to the two-tier encoding considered in this paper. In this encoding, approximately half of modules (i.e.~the X and Z modules) are ancillaries for stabiliser measurements. Without the overhead cost due to X and Z modules, and if physical-qubit stabiliser measurements are continuously performed across the whole module network, isolated two-dimensional error-correction lattices of entanglement errors merge into a single connected lattice. The error correction on the connected lattice, which is essentially three-dimensional, will be harder than the error correction on isolated lattices. However, this disadvantage may be tolerable for very large modules, in which case the overhead cost due to X and Z modules may not be necessary. It would be interesting to perform an analysis of this case where modules are very large, exceeding the size required for storing logical qubits, in order to determine whether the hierarchical code introduced in this paper remains useful in that domain. 

In our analysis, we contrasted the performance of a network of substantial modules with that of a network of small modules, each containing one data qubit of a simple surface code. In terms of the total number of physical qubits needed to achieve a given low error rate at the logical level, our somewhat surprising conclusion was that simple modules are only marginally inferior to large modules containing nearly a thousand qubits. In this sense, our result is that {\it the granularity of a network does not strongly influence the resource costs}. Thus experimentalists are free to build systems with whatever module size suits their particular technologies without paying a significant penalty in total qubit count. However as a caveat we must remark that in our study we have assumed that physical qubits have a long memory time, so that memory errors are negligible on the timescale required to perform the entanglement purification (see Appendix~\ref{sec:AppTime}). If this is not the case, then a significant advantage for large modules could emerge because less purification tiers are necessary. An analysis of this scenario would open the way to a full audit of the time cost of network quantum computing, where the time needed for the multiple rounds of stabilisation in the hierarchical picture is contrasted with the time needed for deep purification circuits. 

\section{Conclusions}

In this paper, we have introduced a variant of the surface code approach to fault tolerant quantum information processing. Our variant is intended to support the network paradigm for quantum computing: the machine is divided into many modules which are connected by noisy inter-links, and each module contains a plurality of well-controlled (low noise) qubits. Our approach is a two-tier hierarchical surface code, where the lower tier involves assigning a `patch' of surface code to each physical module. Errors then occur primarily at patch boundaries and this reduces the dimensionality of the syndrome matching task, thus boosting the threshold. We consider a general scenario where errors occurring on the inter-module links are first purified by broker units and then handled by the hierarchical code. 

Both analytical reasoning and numerical results show that larger modules have advantages: the threshold is higher and the qubit cost is lower. However, the advantage in qubit cost is not significant enough to conclude that large modules are preferred platforms of quantum computing, taking into account the difficulty of building large modules. For small modules, we find that the size $\sim 8$ qubits per module is optimal for practical entanglement noise purification, and the total qubit cost per logical qubit is only fifteen times larger than the cost on a monolithic architecture. Broadly our conclusion is that the granularity of a network-based quantum computer does not strongly affect the total resource costs, with the consequence that experimental efforts can target whatever module size is most convenient for the particular technology. 

\begin{acknowledgments}
This work was supported by the EPSRC platform grant `Molecular Quantum Devices' (EP/J015067/1), and the EPSRC National Quantum Technology Hub in Networked Quantum Information Processing. The authors would like to acknowledge the use of the University of Oxford Advanced Research Computing (ARC) facility in carrying out this work. http://dx.doi.org/10.5281/zenodo.22558.
\end{acknowledgments}

%%%%%%%%%%%%%%%%%%%%%%%%%%%%%
\newpage
\widetext

\appendix

\section{Circuits}
\label{sec:AppCircuits}

\begin{figure}[h]
\centering
\includegraphics[width=0.75\linewidth]{\figpath 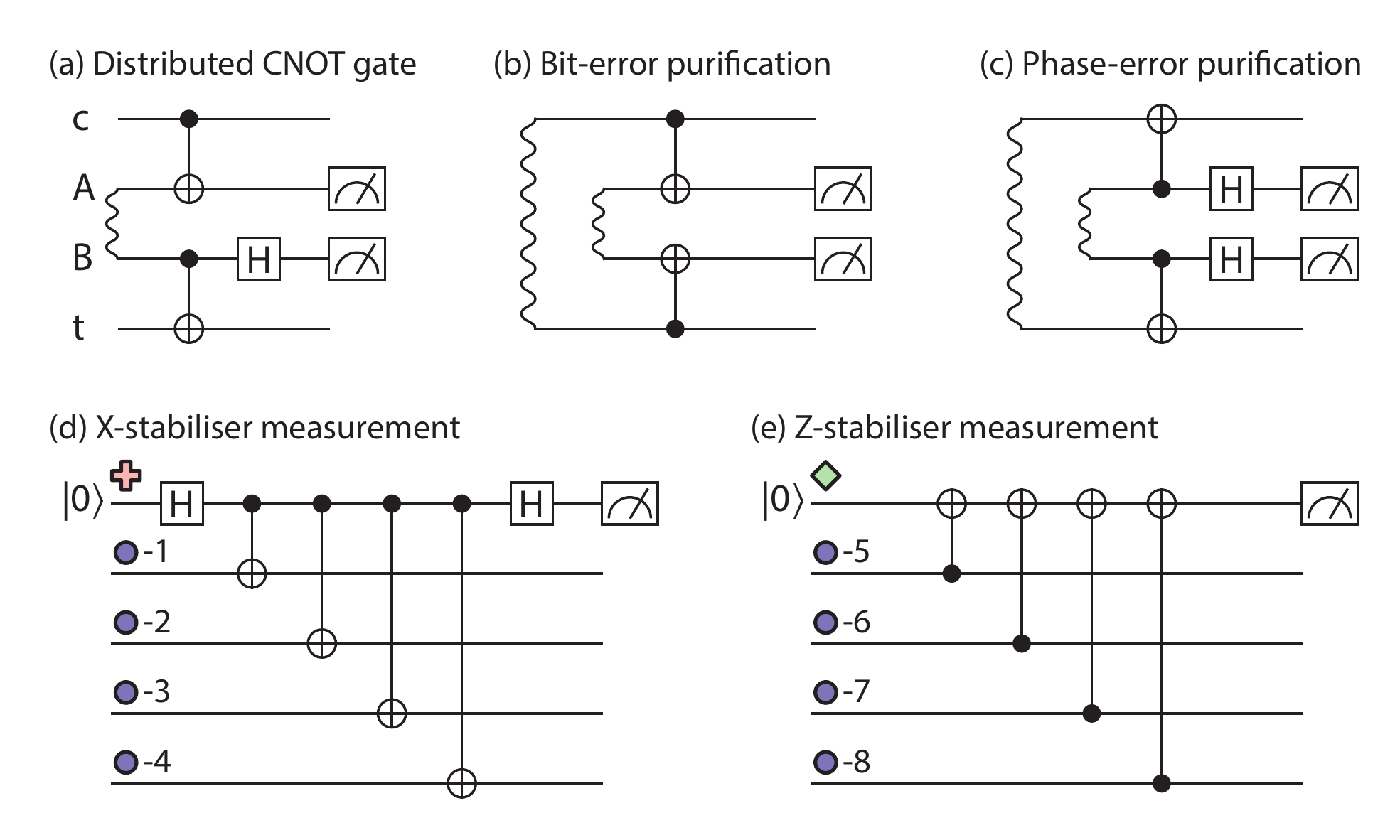}
\caption{
Circuits for the distributed CNOT gate, entanglement purifications and stabiliser measurements. (a) This circuit is equivalent to a CNOT gate on the qubit-c (control) and the qubit-t (target), up to Pauli gates $Z_\text{c}$ and $X_\text{t}$ depending on measurement outcomes. (b) and (c) If the ideal entangled state is in the form $(\ket{00} + \ket{11})/\sqrt{2}$, the output entanglement is discarded if two measurement outcomes are different. In the bit-error purification, the bit-error rate is reduced from $q_\text{B}$ for input entanglement to $\sim q_\text{B}^2$ for the post-selected output entanglement, but the phase-error rate is increased from $q_\text{P}$ to $\sim 2q_\text{P}$. It is similar for the phase-error purification. (d) and (e) Each full round of stabiliser measurements involves both X-stabiliser measurements and Z-stabiliser measurements. Labels of data qubits (see Fig.~\ref{fig:code}) indicate the sequence of CNOT gates (CNOT gates with the same orientation are performed in parallel). 
}
\label{fig:circuit}
\end{figure}

Circuits for performing the distributed CNOT gate, entanglement purifications and stabiliser measurements, are shown in Fig.~\ref{fig:circuit}. 

\section{Module-qubit stabiliser measurements}
\label{sec:AppMQSM}

\begin{figure}[tbp]
\centering
\includegraphics[width=0.7\linewidth]{\figpath 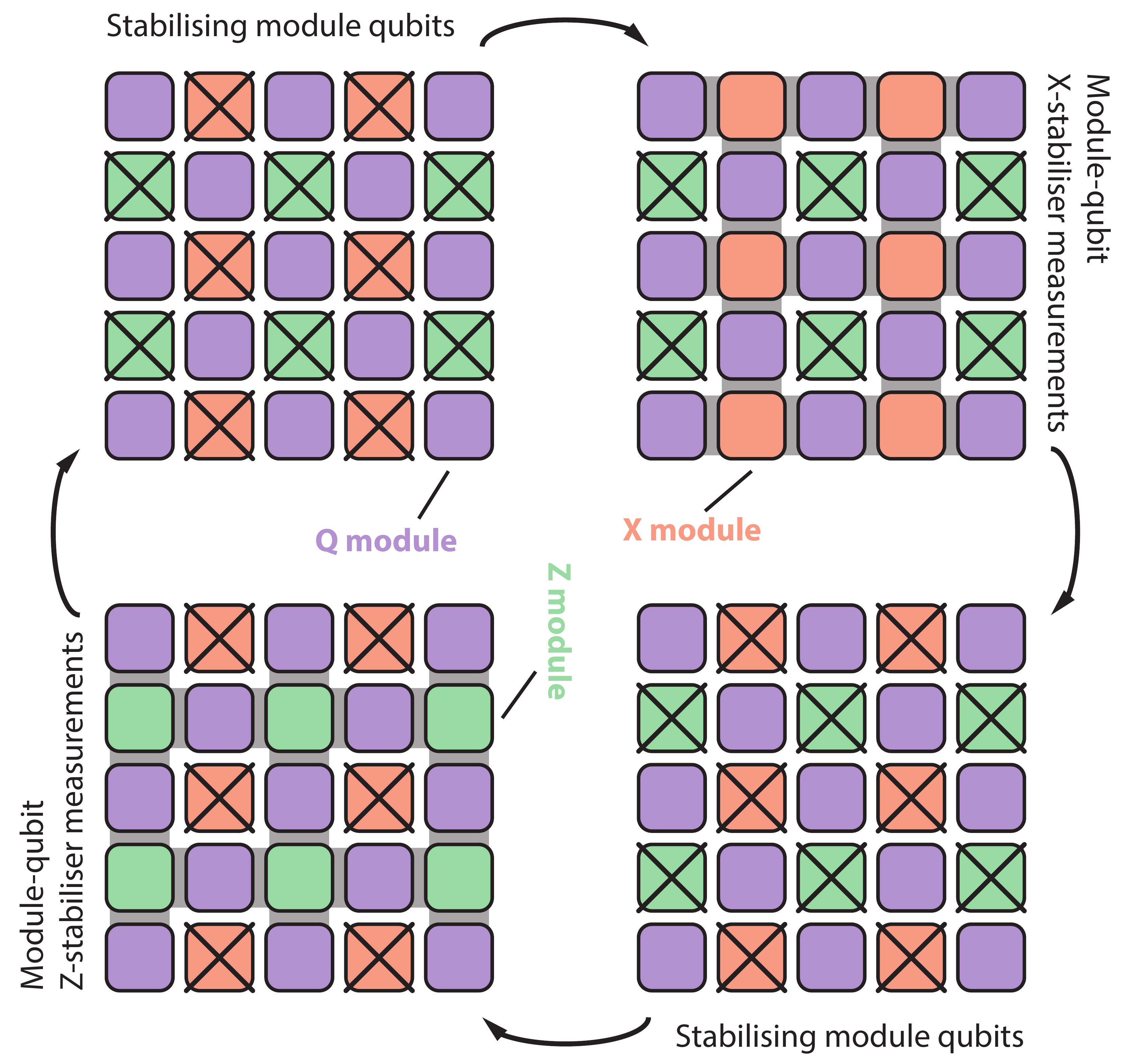}
\caption{
Layout of module-qubit stabiliser measurements. When each module qubit is individually stabilised, only Q modules are involved, and inter-module entanglement is not required. When module-qubit stabilisers are measured, either X modules or Z modules are used to read stabilisers of module qubits, which needs inter-module entanglement for implementing distributed CNOT gates. 
}
\label{fig:flowII}
\end{figure}

The layout of module-qubit stabiliser measurements is shown in Fig.~\ref{fig:flowII}. 

\section{Simulation of thresholds}
\label{sec:AppThreshold}

\begin{figure}[tbp]
\centering
\includegraphics[width=0.65\linewidth]{\figpath 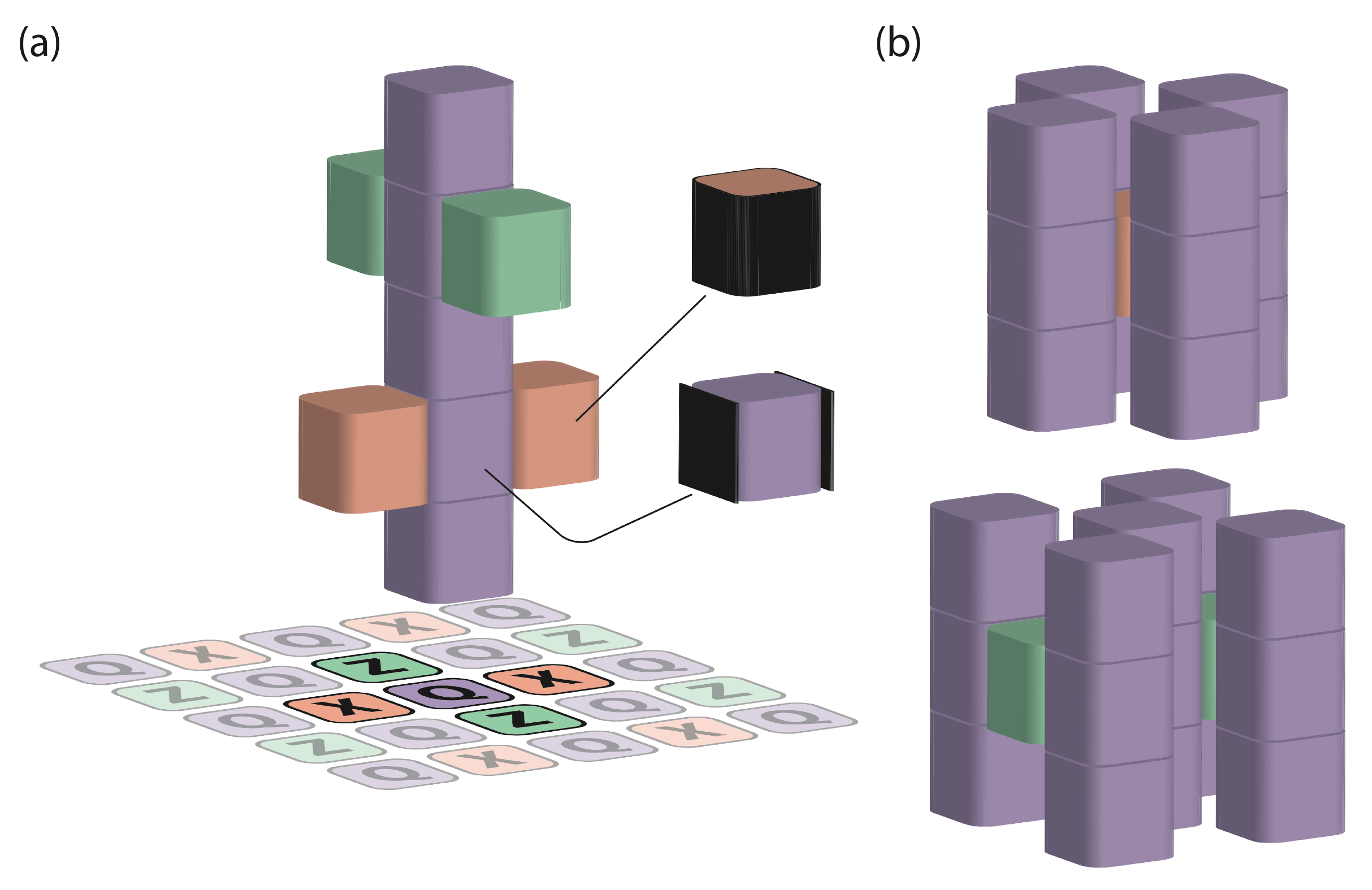}
\caption{
(a) Error correction lattice of physical-qubit errors. When intra-module operations are ideal, measurement errors of module-qubit X stabilisers are all due to errors on the surface of X-module (red) cubes, and module-qubit phase errors are all due to errors on the surface of Q-module (purple) cubes connected with Z-module (green) cubes. It is similar for measurement errors of module-qubit Z stabilisers and module-qubit bit errors. (b) Triple-size lattices. On the triple-size lattice with a X-module cube at the centre, errors near the boundary of the X-module cube are sufficiently considered. On the triple-size lattice with a Q-module cube at the centre, errors near the boundary of the Q-module cube are sufficiently considered.
}
\label{fig:lattice}
\end{figure}

Logical-qubit error rates are obtained using the Monte Carlo method by simulating errors occurring on a logical qubit encoded in a $(2L-1)\times (2L-1)$ module array during $L$ rounds of module-qubit stabiliser measurements. We have used the Edmonds's minimum weight matching algorithm~\cite{Kolmogorov2009} in the surface-code error correction. In our simulations, we have set $n = (D+1)/2$ (see Fig.~\ref{fig:flow}). For simple modules, the error correction is directly performed on a conventional surface code error correction lattice. For large modules ($D \geq 5$), the error correction has two steps. Firstly, physical-qubit errors are corrected on the error correction lattice shown in Fig.~\ref{fig:lattice}(a), which is a three-dimensional lattice formed by cubes with the side length $\sim n$. After the first step, there are only module-qubit errors left, which are further corrected on a conventional surface code error correction lattice representing the array of module qubits.

When intra-module operations are ideal, i.e.~$\epsilon = 0$ (see Fig.~\ref{fig:threshold}), errors (due to entanglement errors) only occur on boundaries of connected cubes [see Fig.~\ref{fig:lattice}(a)]. These boundary surfaces are separated, hence errors on each boundary surface can be individually corrected and simulated. By simulating errors on the surface of a X-module cube [black surface of the red cube in Fig.~\ref{fig:lattice}(a)], we can find the rate $P_\text{M}$ of measurement errors of module-qubit X stabilisers (measurement errors after the physical-qubit error correction). By simulating errors on the surface of a Q-module cube connected with a Z-module cube [black surface of the purple cube in Fig.~\ref{fig:lattice}(a)], we can find the rate $P_\text{P}$ of module-qubit phase errors during one round of module-qubit stabiliser measurements. It is similar for measurement errors of module-qubit Z stabilisers and module-qubit bit errors. With error rates $P_\text{M}$ and $P_\text{P}$, we can simulate module-qubit errors on the conventional surface code error correction lattice to find the rate of logical-qubit errors.

\begin{figure}[tbp]
\centering
\includegraphics[width=0.4\linewidth]{\figpath 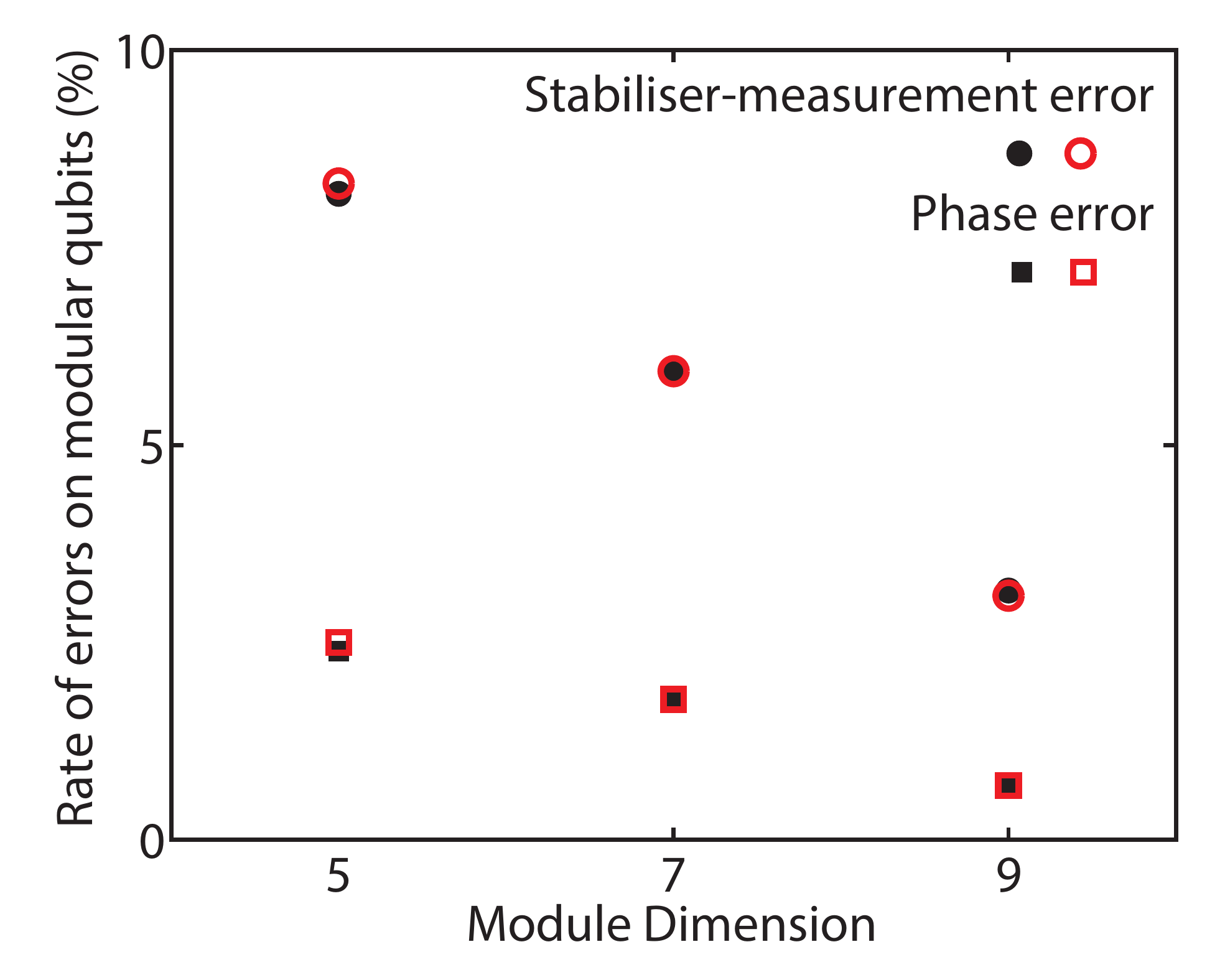}
\caption{
Module-qubit error rates. The rate of measurement errors of module-qubit X stabilisers $P_\text{M}$ and the rate of module-qubit phase errors $P_\text{P}$ are obtained with individual cubes (black solid marks) and triple-size lattices (red empty marks). We have assumed that the (purified) entanglement fidelity $F = 97\%$, entanglement errors are unpolarised, i.e.~$p_\text{X}=p_\text{Y}=p_\text{Z}$, and all intra-module operations have the same error rate $\epsilon_\text{I} = \epsilon_\text{M} = \epsilon_1 = \epsilon_2 = 0.1\%$.
}
\label{fig:triple}
\end{figure}

When intra-module operations are not ideal, e.g.~$\epsilon = 0.1\%$ (see Fig.~\ref{fig:threshold}), errors are not restricted to boundary surfaces. In this case, the first step of the error correction (correcting physical-qubit errors) must be performed on the entire lattice [see Fig.~\ref{fig:lattice}(a)]. However, for large modules ($D \geq 5$), the entire lattice is too large to be directly simulated. Therefore, we individually simulate errors in each cube to approximately calculate the rate of module-qubit errors. We note that, when intra-module operations are not ideal, module-qubit errors also occur when module qubits are stabilised (corresponding to Q-module cubes without contact with X-module or Z-module cubes), which must be taken into account. Because we are interested in the case that intra-module operations have the error rate $\epsilon = 0.1\%$, which is much smaller than $1\%$ the error-rate threshold for intra-module operations~\cite{Wang2011}, probabilities of error chains (after the physical-qubit error correction) decrease rapidly with their lengths. Therefore, by simulating cubes individually, only the effect of errors (induced by intra-module operations) near boundary surfaces are approximately considered. In Fig.~\ref{fig:triple}, we compare module-qubit error rates obtained with individual cubes and error rates obtained on triple-size lattices [see Fig.~\ref{fig:lattice}(b)]. Each triple-size lattice has the dimension $\sim 3n$, in which the boundary effect has been sufficiently considered. Even in the case of the smallest cube ($D = 5$), neglecting the boundary effect only slightly changes error rates of module qubits.

\section{Simulation of qubit costs}
\label{sec:AppCosts}

\begin{figure}[tbp]
\centering
\includegraphics[width=0.8\linewidth]{\figpath 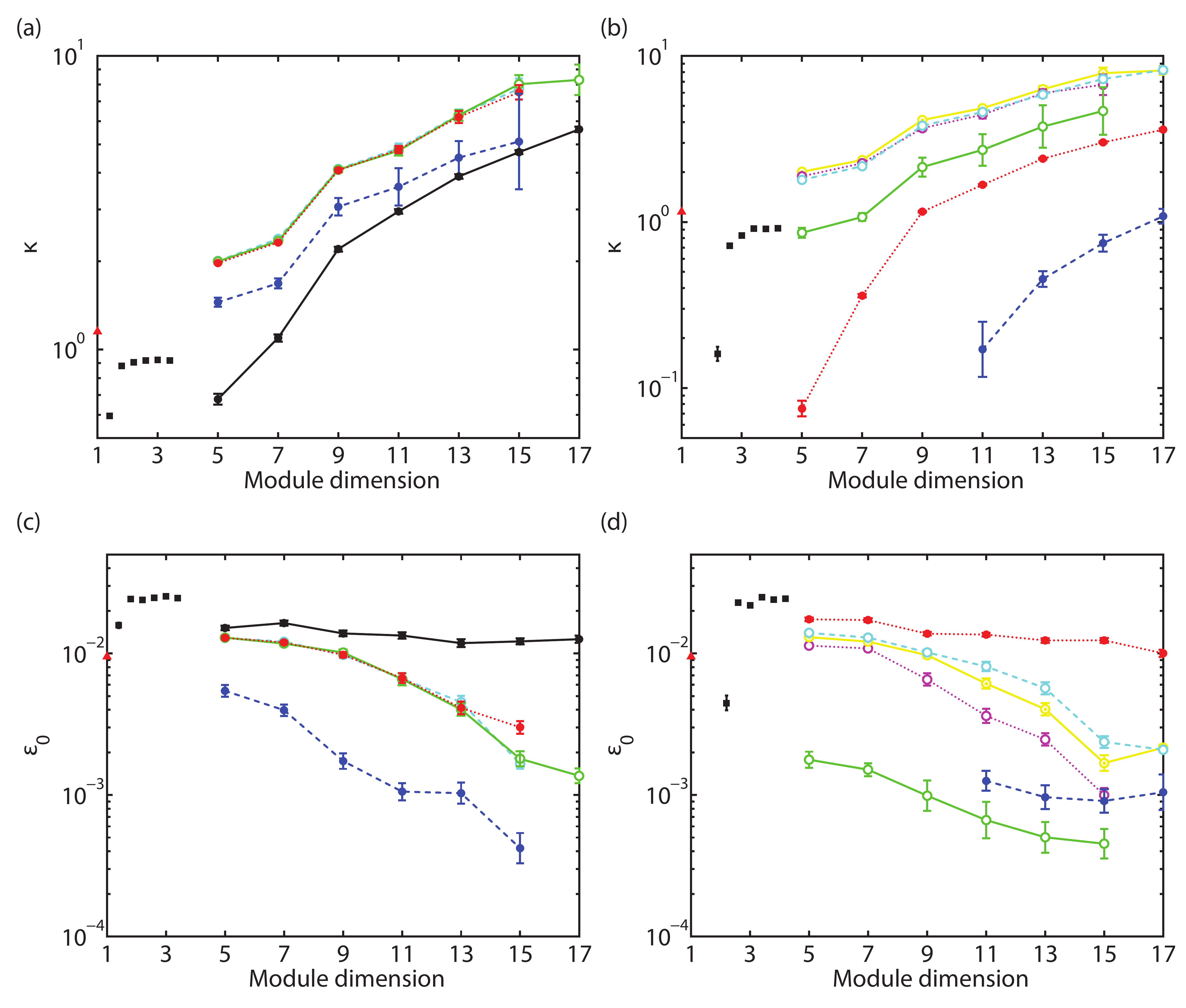}
\caption{
Parameters $\epsilon_0$ and $\kappa$ for entanglement error rates $1.5\%$ [(a) and (c)] and $15\%$ [(b) and (d)]. Marks consist with Fig.~\ref{fig:cost}. For simple modules, the module dimension is always $D = 1$, and the number of purification tiers increases form left to right.
}
\label{fig:parameters}
\end{figure}

The qubit cost is calculated with parameters $\epsilon_0$ and $\kappa$. For a given logical error rate $\epsilon_\text{L}$, one can find the minimum $L$ satisfying $\epsilon_\text{L} \geq \epsilon_0 e^{-\kappa L}$. This minimum $L$ determines the size of the logical qubit. The total number of qubits in each logical qubit is $(2L-1)^2 \times S$, where $S$ is the number of qubits in each module (module size). 

Parameters $\epsilon_0$ and $\kappa$ are shown in Fig.~\ref{fig:parameters}. For simple modules, parameters $\epsilon_0$ and $\kappa$ are obtained by directly fitting logical qubit error rates for $L = 3,5,7,9,11$ with the function
\begin{eqnarray}
\epsilon_\text{L} = \epsilon_0 e^{-\kappa L}.
\end{eqnarray}
For large modules with $D \geq 5$, because module-qubit error rates are very small, it is hard to directly find logical-qubit error rates in simulations. Therefore, we firstly obtain the module-qubit error rates $(P_\text{M},P_\text{P})$, and then we find logical-qubit error rates for module-qubit error rates $(rP_\text{M},rP_\text{P})$. Here, the ratio $r$ is chosen so that $(rP_\text{M},rP_\text{P})$ are large enough for simulating logical-qubit error rates (not far below the threshold). Then, we fit logical-qubit error rates for $L = 3,5,7,9,11$ with the function~\cite{Fowler2012}
\begin{eqnarray}
\epsilon_\text{L} = e^{ (\alpha \ln r + \beta) L + \gamma }.
\end{eqnarray}
With fitting parameters $\alpha$, $\beta$ and $\gamma$, we can obtain parameters $\kappa = \beta$ and $\epsilon_0 = e^\gamma$.

\section{purification and time cost}
\label{sec:AppTime}

\begin{figure}[tbp]
\centering
\includegraphics[width=0.4\linewidth]{\figpath 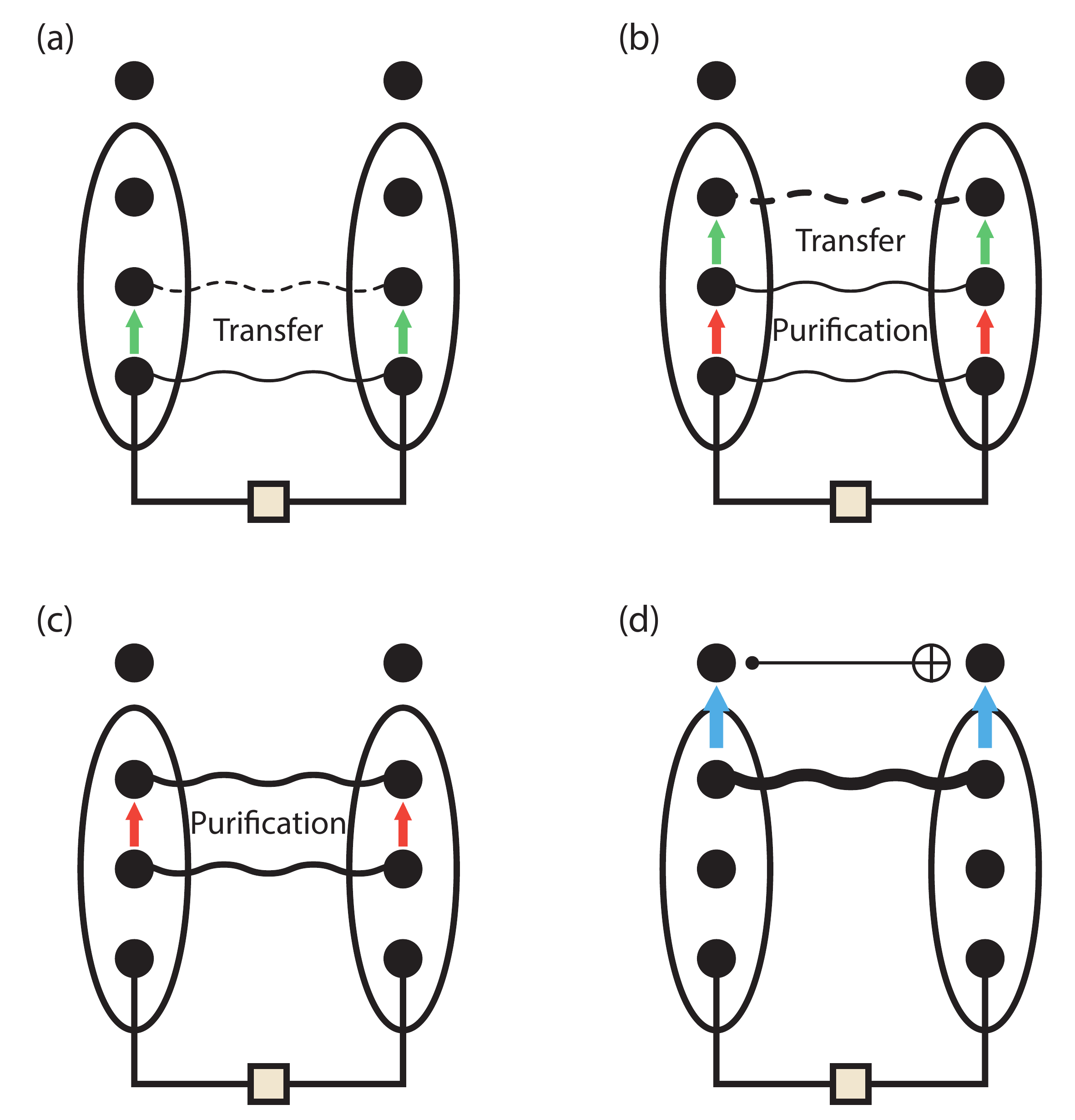}
\caption{
Protocol of the entanglement purification. (a) Firstly, the raw entanglement is generated with optically-coupled qubits (qubits at the bottom), and this raw entanglement is transferred to the upper pair of qubits via swap gates. Each swap gate is realised with three CNOT gates in our simulations. (b) The second raw engagement is generated with optically-coupled qubits and is used to purity the raw entanglement on the upper pair of qubits. The first-tier purified entanglement is then transferred upward. (c) The second first-tier purified entanglement is prepared and used to purify the previous first-tier purified entanglement. With more qubits in each broker, by transferring purified entanglement upward, this purification process continues until reaching the top of the broker. (d) The finally purified entanglement is used to perform the distributed CNOT gates on client qubits.
}
\label{fig:purification}
\end{figure}

In our simulations, we have considered the purification protocol proposed in Ref.~\cite{Jiang2007}, in which phase errors and bit errors are corrected alternatively, i.e.~in each tier of the purification either only phases errors or only bit errors are corrected. Circuits for bit-error purification and phase-error purification are shown in Fig.~\ref{fig:circuit}~(b)~and~(c). The overall purification protocol is shown in Fig.~\ref{fig:purification}. Note that in our simulations, for distributed CNOT gates involving an X (Z) ancillary qubits, phase (bit) errors are corrected in the first tier. 

\begin{table}[h]
\begin{center}
\begin{tabular}{|c|c||c|c|}
\hline
\multicolumn{4}{|c|}{$F = 98.5\%$}																								\\ \hline
\multicolumn{2}{|c||}{$P_\text{S} = 99\%$}	& \multicolumn{2}{c|}{$P_\text{S} = 99.9\%$}		\\ \hline \hline
$n_\text{D}$			& $N$								& $n_\text{D}$			& $N$									\\ \hline
$0$							& $1$								& $0$							& $1$									\\ \hline
$1$							& $4$								& $1$							& $4$									\\ \hline
$2$							& $8$								& $2$							& $12$								\\ \hline
$3$							& $16$							& $3$							& $20$								\\ \hline
$4$							& $32$							& $4$							& $40$								\\ \hline
$5$							& $64$							& $5$							& $80$								\\ \hline
$6$							& $132$							& $6$							& $158$								\\ \hline
$7$							& $268$							& $7$							& $320$								\\ \hline
$8$							& $544$							& $8$							& $646$								\\ \hline
\end{tabular}
\begin{tabular}{|c|c||c|c|}
\hline
\multicolumn{4}{|c|}{$F = 85\%$}																									\\ \hline
\multicolumn{2}{|c||}{$P_\text{S} = 99\%$}	& \multicolumn{2}{c|}{$P_\text{S} = 99.9\%$}		\\ \hline \hline
$n_\text{D}$			& $N$								& $n_\text{D}$			& $N$									\\ \hline
$0$							& $1$								& $0$							& $1$									\\ \hline
$1$							& $6$								& $1$							& $10$								\\ \hline
$2$							& $18$							& $2$							& $26$								\\ \hline
$3$							& $32$							& $3$							& $44$								\\ \hline
$4$							& $66$							& $4$							& $92$								\\ \hline
$5$							& $114$							& $5$							& $150$								\\ \hline
$6$							& $218$							& $6$							& $286$								\\ \hline
$7$							& $420$							& $7$							& $542$								\\ \hline
$8$							& $848$							& $8$							& $1064$							\\ \hline
\end{tabular}
\end{center}
\caption{
The number $N$ of raw entanglement pairs required for obtaining one pair of $n_\text{D}$-tier purified entanglement with the success probability $P_\text{S}$. The failure to prepare a purified entangled state results in missing the corresponding stabiliser measurement for one round, i.e.~the stabiliser measurement is not successful in that round, which could be compensated by enlarging the logical qubit. When $P_\text{S} \sim 99.9\%$, we expect that missing a small portion of stabiliser measurements only increases the resource cost slightly. We have assumed that raw entanglement has the fidelity $F$, entanglement errors are unpolarised, i.e.~$p_\text{X}=p_\text{Y}=p_\text{Z}$, and all intra-module operations have the same error rate $\epsilon_\text{I} = \epsilon_\text{M} = \epsilon_1 = \epsilon_2 = 0.1\%$.
}
\label{tab:number} 
\end{table}

If the time cost of generating raw inter-module entanglement is much higher than the time cost of intra-module operations, the time cost of one round of module-qubit stabiliser measurements is $2n\times N\times \tau$, where half of the $4n$ rounds of physical-qubit stabiliser measurements (see Fig.~\ref{fig:flow}) need inter-module entanglement, $N$ is the number of raw entanglement pairs for preparing one pair of purified entanglement (see Table~\ref{tab:number}), and $\tau$ is the time cost of preparing one pair of raw entanglement. We note that we have taken $n = (D+1)/2$ in our simulations. For simple modules, if each module has only one broker, the time cost is amplified by a factor of $2$. The overall time cost of the computing, i.e.~the total rounds of module-qubit stabiliser measurements, depends on the algorithm, which beyond the scope of this paper.

\end{document}